\documentclass[11pt]{report}
\usepackage{csthesis}
\usepackage{amsmath}
\usepackage{amsfonts}
\usepackage{amssymb}

\newtheorem{theorem}{Theorem}[section]
\newtheorem{lemma}[theorem]{Lemma}

\newtheorem{corollary}[theorem]{Corollary}
\newtheorem{conjecture}[theorem]{Conjecture}

\newenvironment{proof}[1][Proof]{\begin{trivlist}
\item[\hskip \labelsep {\bfseries #1}]}{\end{trivlist}}

\makeatletter
\@addtoreset{theorem}{chapter}
\makeatother

\begin{document}
\setlength{\unitlength}{2mm}
\figurespagetrue


%
%


\title{Extremal problems in ordered graphs}
\author{Craig Weidert}
\qualification{B.S., Harvey Mudd College, 2007}
\degree{Master of Science}
\submitdate{Summer 2009}
\copyrightyear{2009}


\chair{Dr.~Valentine Kabanets} 
\signatory{Dr.~Gabor Tardos,
        Senior Supervisor}
\signatory{Dr.~Binay Bhattacharya,
        Supervisor}
\signatory{Dr.~Luis Goddyn,
        SFU Examiner}

\beforepreface


%
%

\prefacesection{Abstract}
In this thesis we consider ordered graphs 
	(that is, graphs with a fixed linear ordering on their vertices).  
We summarize and further investigations 
	on the number of edges an ordered graph may have 
	while avoiding a fixed forbidden ordered graph as a subgraph.  
In particular, we take a step toward confirming a conjecture of Pach and Tardos \cite{PachTardos}
	regarding the number of edges allowed when the forbidden pattern is a tree 
	by establishing an upper bound for a particular ordered graph 
	for which existing techniques have failed.  
We also generalize a theorem of Geneson \cite{Geneson} 
	by establishing an upper bound on the number of edges allowed 
	if the forbidden graphs fit a generalized notion of a matching.  


%
%

\prefacesection{Acknowledgments}
I would like to thank my parents and my supervisor G\'abor Tardos for their help and support.


\lists

\beforetext

\chapter{Introduction}
\section{Definitions}
In this thesis we summarize and further investigations 
	of extremal graph theory in ordered graphs and 0-1 matrices.  
Formally, an \emph{ordered graph} is a three-tuple $G = (U, <_U, E)$ 
	with $(U, <_U)$ being a linearly ordered set and $E$ as a subset of the pairs of $U$.  
The elements of $U = U(G)$ are the \emph{vertices} of $G$
	and the elements of $E = E(G)$ are the \emph{edges} of $G$.  
If the vertices in a set are consecutive with respect to their ordering, 
	we call them an \emph{interval}.  
Forgetting about the linear ordering of $G$, 
	we have the \emph{underlying graph} $(U, E)$ of $G$.  
\emph{Isomorphism} between two graphs is defined as an isomorphism between the underlying graphs 
	with a preservation of the vertex ordering.  
A \emph{subgraph} of $G$ is a subgraph of the underlying graph of $G$ 
	that inherits the vertex ordering of $G$.  
We define the \emph{degree} $d(u) = d_G(u)$ of a vertex $u \in U$ 
	as its degree in the underlying graph.  

We say that an ordered graph $G$ \emph{contains} another ordered graph $G'$ 
	if $G'$ is isomorphic to a subgraph of $G$.  
If $G$ doesn't contain $G'$, then we say that $G$ \emph{avoids} $G'$.  
In this work we are interested in the maximum number of edges 
	that an ordered graph may have while avoiding a certain fixed ordered graph.  
If an ordered graph $P$ has at least one edge, we call $P$ a \emph{pattern} graph.  
We define the extremal function $ex_<(n, P)$ as the maximum number of edges an ordered graph on $n$ vertices may have 
	while still avoiding a pattern graph $P$.  
For convenience, we sometimes call the graph that may contain the pattern graph the ``host'' graph.  
	
The magnitude of $ex_<$ depends heavily on the interval chromatic number, defined as follows. 
For an ordered graph $G$, let the \emph{interval chromatic number} $\chi_<(G)$ be
    the minimum number of intervals the linearly ordered vertex set of $G$ can be partitioned into
    such that there is no edge between two vertices of the same interval.  
Notably, this quantity is easily computable; a simple greedy algorithm yields $\chi_<(G)$. 
This simplicity stands in stark contrast to the NP-complete task of computing the regular chromatic number of a graph. 
Using Corollary~\ref{ErdosPachTardos} 
	of the Erd\H os, Stone, Simonovits theorem \cite{ErdosStone, ErdosSimonovits} (Theorem \ref{ErdosStoneSimonovits}),  
    we can determine the asymptotic behavior of $ex_<(n, P)$ unless $\chi_<(P) = 2$.  

We can express these more difficult ordered graphs $G$ with $\chi_<(G) = 2$ as \emph{ordered bipartite graphs}.  
Ordered bipartite graphs are a five tuple $G = (U, <_U, V, <_V, E)$.  
These are much the same as ordered graphs 
	except that there are two linearly ordered \emph{partite sets} $(U, <_U)$ and $(V, <_V)$ of vertices
	and $E \subseteq U \times V$.  
We may denote the partite sets of these ordered bipartite graphs as $U(G)$ and $V(G)$.  
When we have an ordered graph $G$ with a unique decomposition into two intervals that are independent sets, 
	we may also speak of $G$ as an ordered (bipartite) graph.  

These ordered bipartite graphs with a unique decomposition into two intervals that are independent sets 
	may be represented using a 0-1 matrices 
	(that is, matrices whose entries are all either zeros or ones)
	by using an adjacency matrix with the rows correspding to one partite set and the columns corresponding to the other.  
Indeed, this is the standard notation used in several previous papers in the area (e.g. \cite{FurediHajnal, Geneson, Keszegh, MarcusTardos, Tardos}).  
Throughout this paper, we use the graph representation of ordered bipartite graphs, but this merely means a change in notation.  
Much as before, we say that a matrix $M$ contains another pattern matrix $M'$ 
	if $M$ contains a submatrix of the same dimensions as $M'$ that has a one entry in every place that $M'$ has a one entry.  
If $M$ doesn't contain $M'$, then $M$ avoids $M'$.  
	
One may find it convenient to restrict both the pattern graphs and the host graphs to those with interval chromatic number two.  
In this way, one may work exclusively with matrix or ordered bipartite graph representations.   
Toward this end, given an ordered bipartite pattern graph $P$, 
	we define $ex_2(n, m, P)$ as the maximum number of edges an ordered bipartite graph $G$ 
	with $|U(G)| = n$ and $|V(G)| = m$ may contain while avoiding $P$.  
We use $ex_2(n, P)$ as shorthand for $ex_2(n, n, P)$.  
We may reverse the order of either or both of the partite sets of $P$ (or even swap the partite sets) 
	to obtain an ordered bipartite graph that is equivalent to $P$ with respect to the extremal function.  
As Pach and Tardos show in Theorem~\ref{bipartite relates}, for bipartite patterns $P$ $ex_<(n, P)$ and $ex_2(n, P)$ are very closely related.  

One last related extremal function has to do with convex geometric graphs.  
A \emph{geometric graph} is a graph drawn in the plane with a set of points as its vertex set
	and its edges drawn as straight line segments.  
The drawn edges are not allowed to pass through the vertices except at their endpoints.  
We consider two geometric graphs isomorphic if their underlying graphs are isomorphic and the edge crossings are preserved.  
A geometric graph is \emph{convex} if its vertices are in convex position.  
Convex geometric graphs are purely combinatoric objects 
	because two edges of convex geometric graphs cross if and only if their endpoints alternate in the cyclic order of the vertices.  
Thus, one may think of these graphs as ordered graphs with a cyclic ordering on their vertices.  
Again, we define pattern, subgraph, contains, and avoids for convex geometric graphs as before.  
For a convex geometric pattern graph $P$, 
	we define $ex_\circlearrowright(n, P)$ as the maximum number of edges that a convex geometric graph with $n$ vertices may contain 
	while avoiding $P$.  
For a convex geometric pattern graph $P$, 
	we define the \emph{circular chromatic number} $\chi_\circlearrowright(n, P)$ 
	as the minimum number of colors necessary for a proper coloring of $P$ 
	in which every color class is an interval with respect to this circular ordering.  
In a way similar to the ordered graph case, $ex_\circlearrowright$ turns out to depend heavily on $\chi_\circlearrowright$.  
For a collection of results on convex geometric graphs, see \cite{BrassKarolyiValtr}.  


\section{Results}
In \cite{MarcusTardos}, Marcus and Tardos prove the linearity of $ex_2(n, P)$ for matchings $P$ 
	and use it to settle the Stanley-Wilf conjecture.  
This conjecture concerns containment and avoidance in the context of permutations.  

Let $m \in \mathbb{Z}^+$.  
We call an ordered graph $G$ an \emph{$m$-tuple matching}
    if $V(G) = \{v_1 < \dots < v_{(m+1)k}\}$, 
	and $E(G) = \{v_jv_{k+i+m(\pi(j)-1)}:i \in \{1, \dots, m\}, j \in \{1, \dots, k\}\}$
    where $\pi:\{1, \dots, k\} \rightarrow \{1, \dots, k\}$ is a permutation.  
Based on Geneson's proof in \cite{Geneson}, 
	we show in Theorem~\ref{matching theorem} that for all $m$-tuple matchings $P$, $ex_<(n, P) = O(n)$.  
This in turn implies similar theorems for $ex_2(n, P')$ (previously shown by Geneson) and $ex_\circlearrowright(n, P'')$ 
	where $P'$ and $P''$ are analogs of $m$-tuple matchings in the ordered bipartite graphs and convex geometric graphs respectively.  

Let us define a \emph{minimally non-linear pattern} as a pattern 
	that has a non-linear extremal function
	and whose proper subgraphs all have linear extremal functions.  
Using the linearity of $ex_2(n, P)$ for $m$-tuple matchings $P$ and Keszegh's previous work in \cite{Keszegh}, 
	in Theorem~\ref{inf min non lin} Geneson confirms Keszegh's conjecture 
	concerning the existence of infinitely many minimally non-linear patterns.  

Pach and Tardos formulated the following conjecture in \cite{PachTardos}:
\begin{conjecture}\label{tree conjecture}
If $P$ is a pattern with the underlying graph of $P$ being a forest,
	then $$ex_2(n, P) \leq n(\log n)^{O(1)}.$$
\end{conjecture}
 
\begin{figure}[h!]
\centering
\begin{picture}(12, 8)
\multiput(0,8)(6,0){3}{\circle*{1}} 
\multiput(0,0)(4,0){4}{\circle*{1}} 
\put(0,0){\line(0,1){8}} 
\put(0,0){\line(3,4){6}} 
\put(4,0){\line(1,1){8}} 
\put(8,0){\line(-1,1){8}} 
\put(12,0){\line(-3,4){6}} 
\put(12,0){\line(0,1){8}} 
\end{picture}
\caption{The sailboat graph.}
\label{sailboat}
\end{figure}
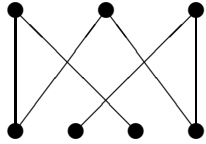

Using these rules outlined in Lemma~\ref{tree rules}, Conjecture \ref{tree conjecture} has been proven for all trees with at most five edges. 
One of the two tree patterns $P$ with six edges for which the conjecture is still open is shown in Figure~\ref{sailboat} 
	where $U(P)$ is set of vertices in the top row, 
	the $V(P)$ is the set in the bottom row, 
	and the linear orderings of the vertices go from left to right.  
We call this graph the ``sailboat graph''.  
In Theorem~\ref{s_matrix theorem}, we prove a bound of $O(n2^{\sqrt{\log n \log\log n}})$ for the extremal function of this graph. 
The logarithms in this theorem and elsewhere are binary.  
This bound is slightly weaker than the bound in Conjecture~\ref{tree conjecture}, but still below $O(n^{1+\epsilon})$ for any $\epsilon$. 
In particular, this bound is below the number of edges allowed in any pattern graph with a $k$-cycle in its underlying graph, 
	shown in Theorem~\ref{cycle theorem} to be $\Omega(n^{1 + 1/(k-1)})$.  
 
\chapter{Previous Work}\label{previous}

\section{Preliminary Results}
The magnitude of $ex_<(n, P)$ depends heavily on the interval chromatic number.  
Using Corollary~\ref{ErdosPachTardos}, 
    we can determine the asymptotic behavior of $ex_<(n, P)$ unless $\chi_<(P) = 2$. 
This corollary follows from the Erd\H os, Stone, Simonovits theorem \cite{ErdosStone, ErdosSimonovits} (Theorem~\ref{ErdosStoneSimonovits}) 
	which concerns $ex(n, P)$, the number of edges that an unordered graph may contain 
	while avoiding an unordered graph $P$.  
   
\begin{theorem}\label{ErdosStoneSimonovits}(Erd\H os, Stone, Simonovits \cite{ErdosStone, ErdosSimonovits})
Let $G$ be a fixed graph.  
Then $$ex(n, K_{\chi(G)}) \leq ex(n, G) \leq ex(n, K_{\chi(G)}) + o(n^2).$$  
\end{theorem}
Here $K_{\chi(G)}$ is the complete graph on $\chi(G)$ vertices 
	and $ex(n, K_{\chi(G)})$ is the number of edges in the Tur\'an graph $T_{n, \chi(G)-1}$.  
The \emph{Tur\'an graph} $T_{n, r}$ is the complete $r$-partite graph with either $\lfloor n/r \rfloor$ or $\lceil n/r \rceil$ vertices in each class.  
Thus, the number of edges in the Tur\'an graph $T_{n, \chi(G) - 1}$ is $\left(1-1/(\chi(G)-1)\right)\binom{n}{2} + O(n)$.  
We omit the proof of this theorem.  

\begin{corollary}\label{ErdosPachTardos}(Pach, Tardos \cite{PachTardos})
The maximum number of edges $ex_<(n, P)$ that an ordered graph with $n$ vertices may have while avoiding another ordered graph $P$ satisfies
$$\left(1-\frac{1}{\chi_<(P) - 1}\right)\binom{n}{2} < ex_<(n, P) = \left(1-\frac{1}{\chi_<(P) - 1}\right)\binom{n}{2} + o(n^2).$$ 
\end{corollary}

\begin{proof}
Let $T^{<}_{n, r}$ be the ordered version of the Tur\'an graph in which the vertices in each class form an interval.  
For the first inequality, notice that $\chi_<\left(T^{<}_{n, \chi(P)-1}\right) = \chi(P)-1$.  
Thus, $T^{<}_{n, \chi(P)-1}$ avoids all graphs $G$ with $\chi(G) \geq \chi(P)$.  
Since $\left|E\left(T^{<}_{n, \chi(P)-1}\right)\right| = \left(1-\frac{1}{\chi_<(P) - 1}\right)\binom{n}{2} + O(n)$, we have the first inequality.  

For the second inequality, we appeal Theorem~\ref{ErdosStoneSimonovits}.  
We may show 
	that for an ordered graph $G$ with $|V(G)| = n > n_0(k, m, \varepsilon)$ 
	with at least $\left(1-\frac{1}{\chi_<(P) - 1}\right)\binom{n}{2} + \varepsilon n^2 \approx \left|E\left(T^{<}_{n, \chi(P)-1}\right)\right| + \varepsilon n^2$ edges, 
	$G$ has $T^{<}_{mk, k}$ as a subgraph 
	and thereby contains all ordered graphs with interval chromatic number at most $k$ and at most $m$ vertices.  
By Theorem~\ref{ErdosStoneSimonovits}, 
	an ordered graph $G$ with $|V(G)| = n > n_0(k, mk^3, \varepsilon)$
	will have $T_{mk^3, k}$ as an underlying subgraph.  

Partition the vertices of our ordered graph $G$ into $k$ intervals of $mk^2$ consecutive vertices each.  
Now, imagine a bipartite graph with the classes of the Tur\'an graph $T_{mk^3, k}$ as one set of vertices 
	and the $k$ intervals of our ordered graph $G$ as the other set of vertices.  
Let this bipartite graph have an edge between a class and an interval if there are at least $m$ vertices from that class in that interval.  
This bipartite graph must have a matching and so $G$ must contain $T^{<}_{mk, m}$ 
	and thereby all ordered graphs with interval chromatic number at most $k$ and at most $m$ vertices.  
(Assume to the contrary that there is no such matching.  
Then by Hall's Theorem, there is some subset of $i$ classes 
	such that there are at most $i-1$ intervals that each have at least $m$ vertices in each of those $i$ classes.  
This means that the other $k-i+1$ intervals have at most $m-1$ vertices in each of those $i$ classes.  
These $k-i+1$ intervals may also collectively have $(k-i)mk^2$ vertices in the other classes.  
However, these $k-i+1$ intervals have a total of $(k-i+1)(mk^2)$ vertices total and this is greater than $i(k-i+1)(m-1)+(k-i)mk^2$.  
So, we have a contradiction and so $G$ must contain $T^{<}_{mk, m}$.)  
$\Box$
\end{proof}

\section{The relationship between $ex_<$ and $ex_2$}
As stated in the introduction, due to Corollary~\ref{ErdosPachTardos} 
	we generally restrict our attention to pattern graphs $P$ with $\chi_<(P) = 2$. 
Note that unless $P$ has isolated vertices, the decomposition of the vertices of $P$ into two independent sets is unique. 
For convenience, we may also restrict our host graphs to those that have interval chromatic number 2.  
As Theorem~\ref{bipartite relates} shows, we do not lose too much by restricting ourselves to host graphs $G$ where $\chi_<(G) = 2$. 
First, we need the following lemma.  

\begin{lemma}(Pach, Tardos \cite{PachTardos})\label{bipartite relates helper} 
\begin{itemize}
\item[(a)] 
For any ordered graph $G$ with $n$ vertices, 
	there exist edge disjoint subgraphs $G_i$ for $0 \leq i \leq \lceil \log n \rceil$
	such that $\cup_{i = 0}^{\lceil \log n \rceil} E(G_i) = E(G)$ 
	and each connected component of $G_i$ has at most $\lceil n/2^i \rceil$ vertices and interval chromatic number at most 2.  
\item[(b)](super-additivity)
For any pattern $P$ and positive integers $m$ and $n$, 
$$ex_2(n+m, P) \geq ex_2(n, P) + ex_2(m, P).$$  
\end{itemize}
\end{lemma}

\begin{proof}
For part (a), denote the vertices of $G$ by $V(G) = \{v_0 < \dots < v_{n-1}\}$ 
	and let $E(G_i) = \{v_jv_k : v_jv_k \in E(G), 
		\lfloor2^ij/n\rfloor = \lfloor2^ik/n\rfloor, 
		\lfloor2^{i+1}j/n\rfloor \neq \lfloor2^{i+1}k/n\rfloor\}$.  
These subgraphs satisfy the requirements of part (a).  

For part (b), we'll show the case when the partite sets of the bipartite ordered graphs are not necessarily the same size.  
Suppose that the first and last vertices of each of the partite sets of $P$ are each incident to at least one edge.  
Label an edge incident to the first vertex in $U(P)$ red and an edge incident to the last vertex in $U(P)$ blue.  
Assume without loss of generality that the vertex of $V(P)$ incident on the red edge 
	is no greater than the vertex of $V(P)$ incident on the blue edge.  
(We can simply use one of the equivalent bipartite ordered graphs if need be.)  
Now say that we have ordered bipartite graphs $A$ and $B$ on disjoint sets of vertices that avoid $P$ 
	with $|U(A)| = n_A$, $|V(A)| = m_A$, $|U(B)| = n_B$, and $|V(B)| = m_B$.  
Let us define an ordered bipartite graph $C$ 
	with $U(C) = U(A) \cup U(B)$, $V(C) = V(A) \cup V(B)$, $E(C) = E(A) \cup E(B)$, 
	and the vertices from $B$ following those from $A$ with respect to the ordering.  
The ordered bipartite graph $C$ avoids $P$.  
Assume to the contrary that $C$ contains $P$.  
On one hand, if the red edge of $P$ is in the edges from $E(B)$, then $B$ would have to contain $P$.  
Similarly, if the blue edge of $P$ is in the edges from $E(A)$, then $A$ would have to contain $P$.  
On the other hand, if the red edge is in the edges from $E(A)$ and the blue edge is in the edges from $E(B)$,  
	then the vertex in $V(C)$ adjacent to the red edge is greater than the vertex in $V(C)$ adjacent to the blue edge.  
This is a contradiction.  

Now, suppose that, for instance, the first vertex of $U(P)$ has no edges incident to it.  
Let $P'$ be ordered bipartite graph obtained by removing the first vertex in $U(P)$ from $P$.  
Let $C$ be any ordered bipartite graph with $|U(C)| = n$ and $|V(C)| = m$ 
	and let $C'$ be the ordered bipartite graph obtained by removing the first vertex of $V(C)$ from $C$.  
Notice that $C'$ avoids $P'$ if an only if $C$ avoids $P$.  
So, in this case $ex_2(n, m, P) = ex_2(n-1, m, P') + m$.  
So, in adding an isolated vertex that is either the largest or smallest of a partite set of a pattern graph we preserve super-additivity.  
So, all pattern graphs are super-additive.  
$\Box$ 
\end{proof}

\begin{theorem}\label{bipartite relates}(Pach, Tardos \cite{PachTardos})
Let $P$ be an ordered graph with a unique decomposition into two intervals that are independent sets.  
Then
	$$ex_2(\lfloor n/2 \rfloor, P) \leq ex_<(n, P) = O(ex_2(n, P)\log n).$$ 
Moreover, if $ex_2(n, P) = O(n^c)$ for some $c>1$, then $ex_<(n, P) = O(n^c)$. 
\end{theorem}

\begin{proof}
The first inequality follows from the fact that any bipartite ordered graph $G$ 
	with $|U(G)| = \lfloor n/2\rfloor$, 
	$|V(G)| = \lfloor n/2\rfloor$, 
	and $|E(G)| = ex_2(\lfloor n/2 \rfloor, P)$ avoiding $P$
	may be translated into an ordered graph which avoids $P$ by simply concatenating the vertex sets.  

For the second inequality, let $G$ be an ordered graph with $n$ vertices that avoids $P$
	and let $G_i$ be subgraphs as in Lemma~\ref{bipartite relates helper}(a).  
Let $C_i$ be the set of connected components of $G_i$.  
Because $G$ avoids $P$, all connected components of these subgraphs $G_i$ avoid $P$ also.  
So, because all of the connected components have interval chromatic number two, 
	for each connected component $g$, $|E(g)| \leq ex_2(|V(g)|, P)$.  
If $ex_2(n, P) = O(n^c)$ for some $c > 1$, then summing over all connected components, we get that $|E(G)| = O(n^c)$.  
Generally, since $ex_2$ is super-additive, we have that for each subgraph $G_i$, 
	$$|E(G_i)| = \sum_{g \in C_i} |E(g)| \leq \sum_{g \in C_i} ex_2(|V(g)|, P) \leq ex_2(n, P).$$  
So, since there are at most $O(\log n)$ subgraphs $G_i$ and they partition the edge set of $G$, $ex_<(n, P) \leq O(ex_2(n, P)\log n)$.  
$\Box$
\end{proof}

As the following example shows, this bound on the loss incurred by restricting attention to host graphs with interval chromatic number 2 is tight. 
Consider the pattern graph $P$ with vertices $\{1, 2, 3, 4\}$ and edges $\{\{1, 4\}, \{1, 3\}, \{2, 4\}\}$. 
Consider an ordered bipartite graph $G$ 
	with $U(G) = \{u_1 < \dots < v_{\lfloor n/2\rfloor}\}$ and $V(G) = \{v_1 < \dots < v_{\lceil n/2\rceil}\}$ that avoids $P$. 
For each vertex of $U$, let us remove the edge that is incident on the smallest vertex of $V$ (if the vertex of $U$ is incident to at least one edge). 
Notice that in this step we remove at most $\lfloor n/2 \rfloor$ edges. 
We are left with a graph that may have at most one edge per vertex in $V$ for a total of at most $\lceil n/2 \rceil$ edges. 
Assume to the contrary that there is a vertex $v$ of $V$ with two edges. 
Then the deleted edge of the smaller neighbor of $v$ along with those two edges incident on $v$
    form the avoided subgraph $P$ which is a contradiction. 
So, if our host graph is restricted to being bipartite, then it may have at most $n$ edges. 
On the other hand, if we allow host graphs $G$ with $\chi_<(G) > 2$, then we may have as many as $\Theta(n \log n)$ edges. 
Consider the host graph $G$ with $V(G) = \{v_1 < \dots < v_n\}$ and $E = \{v_iv_j : |i - j| = 2^k, k \in \mathbb{Z}\}$. 
This graph has $\Theta(n\log n)$ edges and avoids $P$.

\section{The Stanley-Wilf Conjecture}

In \cite{MarcusTardos}, Marcus and Tardos proved the F\"uredi-Hajnal conjecture (Theorem~\ref{Furedi-Hajnal}).  
\begin{theorem}\label{Furedi-Hajnal}(Marcus, Tardos \cite{MarcusTardos})
For ordered bipartite matchings $P$, $ex_2(n, P) = O(n)$.  
\end{theorem} 
This result follows from Theorem~\ref{matching theorem}.  
In \cite{Klazar}, Klazar had previously shown that the F\"uredi-Hajnal conjecture implies 
	the Stanley-Wilf conjecture (Theorem~\ref{Stanley-Wilf}).  
This theorem concerns containment and avoidance in the context of permutations.  
For two permutations $\sigma:\{1, \dots, n\}\rightarrow\{1, \dots, n\}$ and $\pi:\{1, \dots, k\}\rightarrow\{1, \dots, k\}$, 
	we say that $\sigma$ contains $\pi$ 
	if there exist integers $1 \leq x_1 < \dots < x_k \leq n$ 
	such that for $1 \leq i < j \leq k$, 
	$\sigma(x_i) < \sigma(x_j)$ if and only if $\pi(i) < \pi(j)$.  
If $\sigma$ does not contain $\pi$, then $\sigma$ avoids $\pi$.  
For any permutation $\pi:\{1, \dots, k\}\rightarrow\{1, \dots, k\}$ 
	define its corresponding ordered bipartite matching $graph(\pi)$ as 
	$U(graph(\pi)) = \{u_1 < \dots < u_k\}$, 
	$V(graph(\pi)) = \{v_1 < \dots < v_k\}$, 
	and $E(graph(\pi)) = \{u_iv_{\pi(i)} : i \in \{1, \dots, k\}\}$.)  

\begin{theorem}\label{Stanley-Wilf}(Marcus, Tardos \cite{MarcusTardos})
Let $S(n, \pi)$ denote the number of $n$-permutations avoiding $\pi$.  
For all permutations $\pi$, there exists a constant $c = c_\pi$ such that $S(n, \pi) \leq c^n$.  
\end{theorem}

\begin{proof}
This proof is taken from \cite{MarcusTardos} which in turn summarized it from Klazar's argument in \cite{Klazar}.  
For a graph $P$, let $T(n, P)$ be the set of all graphs $G$ avoiding $P$ with $|U(G)| = |V(G)| = n$.  
Since each permutation $\sigma$ that avoids a fixed permutation $\pi$ 
	has a corresponding ordered bipartite graph $graph(\sigma)$ that avoids $graph(\pi)$, 
	$|T(n, graph(\pi))| \geq S(n, \pi)$.  
We now prove that for every permutation $\pi$, there is a constant $c = c_\pi$ such that $|T(n, graph(\pi))| \leq c^n$.  
This implies the theorem.  

Using Theorem~\ref{matching theorem}, we see that for every permutation $\pi$, $ex_2(n, graph(\pi)) = O(n)$.  
We can set up the simple recursion $$|T(2n, graph(\pi))| \leq |T(n, graph(\pi))|15^{ex_2(n, graph(\pi))}.$$
We can map each graph $G$ in $T(2n, graph(\pi))$ into $T(n, graph(\pi))$ 
	by partitioning both $U(G)$ and $V(G)$ into consecutive two-vertex intervals, 
	collapsing these intervals into single vertices, 
	and putting an edge between two vertices in the new graph if there is an edge between the vertices of their corresponding intervals in $G$.  
This new graph also avoids $graph(\pi)$.  
Notice that each graph in $T(n, graph(\pi))$ is the image of at most $15^{ex_2(n, graph(\pi))}$ graphs of $T(2n, graph(\pi))$ 
	since each graph in $T(n, graph(\pi))$ has at most $ex_2(n, graph(\pi))$ edges.  
Because by Theorem~\ref{Furedi-Hajnal} $ex_2(n, graph(\pi) = O(n)$, 
	we can solve this recursion to get $T(n, graph(\pi)) = O(c^n)$ for some constant $c$ depending on $\pi$.  
So, the theorem holds.  
$\Box$
\end{proof}

\section{$m$-tuple matchings and minimally non-linear patterns}
In \cite{Geneson}, Geneson defines what we call $m$-tuple bipartite matchings.  
These are similar to $m$-tuple matchings except that there are two partite vertex sets.  
Formally, an ordered bipartite graph $G$ is a \emph{$m$-tuple bipartite matching}
    if $U(G) = \{u_1 < \dots < u_k\}$, 
	$V(G) = \{v_1 < \dots < v_{mk}\}$, 
	and $E(G) = \{(u_j, i+m(\pi(j)-1)) : i \in \{1, \dots, m\}, j \in \{1, \dots, k\}\}$
    where $\pi:\{1, \dots, k\} \rightarrow \{1, \dots, k\}$ is a permutation.  
\begin{theorem}\label{geneson main}(Geneson, \cite{Geneson})
If $P$ is an $m$-tuple bipartite matching, then $ex_2(n, P) = O(n)$.  
\end{theorem}
This theorem follows from Theorem~\ref{matching theorem} and is proven in the next section.  

In \cite{Keszegh}, Keszegh defines the notion of \emph{minimally non-linear} for ordered bipartite graphs.  
By minimally non-linear, we mean a graph that has a non-linear extremal function 
	but whose proper subgraphs all have linear extremal functions.  
Interestingly, a graph has a linear extremal function if and only if it avoids all minimal non-linear patterns.  
He goes on to conjecture that there are infinitely many minimally non-linear patterns
	and defines a collection $\mathcal{H} = \{H_k\}$ of graphs that he believes have this property.  
For $k \geq 1$, he defines the bipartite ordered graph $H_k$ as 
$U(H_k) = \{u_1 < \dots < u_{3k+4}\}$, 
$V(H_k) = \{v_1 < \dots < v_{3k+4}\}$, 
$E(H_k) = \{u_4v_1, u_1v_2, u_1v_3, u_{3k+3}v_{3k+4}, u_{3k+2}v_{3k+4}\} 
			\cup \{u_{3i+4}v_{3i+1}, u_{3i-1}v_{3i+3}, u_{3i}v_{3i+2} : 1 \leq i \leq k\}$

Using his theorem about $m$-tuple matchings, 
	Geneson in \cite{Geneson} confirms Keszegh's conjecture using this collection of graphs.  
Rather than prove directly that the graphs in this collection are all minimally non-linear, 
	he shows that $H_k$ contains a minimal non-linear pattern large enough to imply the conjecture.  
First we must show that the patterns are non-linear at all.  

\begin{lemma}(Keszegh \cite{Keszegh}) 
For $k \geq 1$, $ex_2(n, H_k) = \Omega(n \log n)$.  
\end{lemma}

\begin{proof}\label{non lin helper}
For this proof we require a new matrix to avoid our matrices $H_k$.  
We present the graph $G$ 
	with $U(G) = \{u_1 < \dots < u_n\}$, $V(G) = \{v_1 < \dots < v_n\}$, 
	and $E(G) = \{u_iv_j : j-i = 3^k, k\in \mathbb{Z}\}$.  
This graph has $\Theta(n \log n)$ edges and avoids $H_k$ for all $k \geq 0$.  

First, notice that for vertices $u_{i_1} \leq u_{i_2} < u_{i_3} < u_{i_4} < u_{i_5} \in U(G)$ 
	and $v_{i_1} < v_{i_2} < v_{i_3} < v_{i_4} \leq v_{i_5} \in V(G)$ 
	with edges $u_{i_1}v_{j_3}, u_{i_2}v_{j_2}, u_{i_3}v_{j_5}, u_{i_4}j_{i_4}, u_{i_5}j_{i_1} \in E(G)$, 
	we have that $(j_3-j_2) - (i_2-i_1) < (j_5 - j_4) - (i_4 - i_3)$.  
We have that $j_1-i_5 = 3^{k_1}, j_2-i_2 = 3^{k_2}, j_3-i_1=3^{k_3}, j_4-i_4=3^{k_4}$, and $j_5-i_3=3^{k_5}$ for $k_3 > k_2$ and $k_5 > k_4$.  
So, we have that $(j_5-j_4) + (i_4-i_3) = 3^{k_5}-3^{k_4} > 2\cdot3^{k_4} \geq j_3-j_2$.  
Also, $(j_3-j_2)+(i_2-i_1) > 3^{k_3}-3^{k_2} \geq 2\cdot3^{k_2} \geq i_4 - i_3$.  
Summing this two inequalities yields $(j_3-j_2) - (i_2-i_1) < (j_5 - j_4) - (i_4 - i_3)$.  

Now, assume to the contrary that $G$ contains $H_k$ for some $|E(G)| = n$.  
Let $G'$ be the subgraph representing $H_k$ with vertices 
	$U(G') = \{u'_{i_1} < \dots < u'_{i_m}\}$
	and $V(G') = \{v'_{j_1} < \dots < v'_{j_m}\}$
	where $m = 3k+4$.  
For $0 \leq l \leq k+1$, 
	let $x_l = j_{3l+3} - j_{3l_2} - i_{3l} + i_{3l-1}$ 
	(define $u'_{i_{-1}} = u'_{i_0} = u'_{i_1}$ and $v'_{j_{m+1}} = v'_{j_{m+2}} = v'_{j_m}$).  
Now, for $0 \leq l \leq k$, 
	observe indices $i_{3l-1} \leq i_{3l} < i_{3l+2} < i_{3l + 3} < i_{3l+4}$
	and $j_{3l+1} < j_{3l+2} < j_{3l+3} < j_{3l+5} \leq j_{3l+6}$
	and notice that since $G$ contains $H_k$, we have the edges required in the above paragraph.  
So, by the above paragraph, we can deduce that $x_l < x_{l+1}$.  
However, $x_0 = j_3 - j_2 > 0$ and $x_{k+1} = i_{m-2} - i_{m-1} < 0$.  
This is a contradiction and so $G$ must avoid $H_k$.  
\end{proof}

Now we show that there are linear patterns embedded within the patterns of $\mathcal{H}$ 
	that are large enough to prove the conjecture.  
From Lemma~\ref{inf min non lin helper}, we will be able to prove the existence of infinitely many minimally non-linear patterns.  

\begin{lemma}\label{inf min non lin helper}(Geneson \cite{Geneson})
For $k \geq 1$, there are at least $k+5$ edges in $H_k$ 
	such that removing any one of them results in a pattern with a linear extremal function.  
\end{lemma}

\begin{proof}
Let $k \geq 1$ and $H_k$ defined as above.  
We will give three categories of edges 
	such that removing an edge from any one of them results in an ordered bipartite graph with a linear extremal function.  
The first category contains the two edges $u_4v_1$ and $u_{3k+4}v_{3k+1}$.  
Removing the edge $u_4v_1$ results in a subgraph of a 2-tuple matching 
	with a single vertex appended to the end of a partite set whose single neighbor 
	is also adjacent to the previously-last vertex of that partite set.  
By Theorem~\ref{geneson main} and Lemma~\ref{tree rules}(b), such a graph has a linear extremal function.  
By symmetry, removing edge $u_{3k+4}v_{3k+1}$ also results in a ordered bipartite graph with a linear extremal function.  

The second category contains the edges $u_1v_2$, $u_1v_3$, $u_{3k+2}v_{3k+4}$, and $u_{3k+3}v_{3k+4}$.  
Removing any of these four edges results in an ordered bipartite graph that is equivalent to a subgraph of a 2-tuple matching.  
Thus, such a graph would have a linear extremal function.  

The last category of edges contains all edges of the form $u_{3i+4}v_{3i+1}$ for $1 \leq i \leq k-1$.  
Removing any of these edges results in a graph that can be decomposed into two graphs $P$ and $Q$
	where $P$ is the induced subgraph on vertices $u_1, \dots, u_{3i+3}$ and $v_1, \dots, v_{3i+3}$
	and $Q$ is the induced subgraph on vertices $u_{3i+4}, \dots, u_{3k+4}$ and $v_{3i+4}, \dots, v_{3k+4}$.  
Let $P'$ be the graph obtained from $P$ by appending a single vertex to the ends of each of the partite sets 
	and creating an edge between these two vertices.  
Similarly, let $Q'$ be the graph obtained from $Q$ by appending a single vertex to the beginnings of each of the partite sets
	and creating an edge between these two vertices.  
Notice that both $P'$ and $Q'$ are subgraphs of (possibly equivalent versions of) 2-tuple matchings and so have linear extremal functions.  
These two ordered bipartite graphs may be combined using Lemma~\ref{tree rules}(d) 
	to create the graph which contains the graph obtained by removing edge $u_{3i+4}v_{3i+1}$.  
By Lemma~\ref{tree rules}(d) this graph, and thus the graph obtained by removing edge $u_{3i+4}v_{3i+1}$, have linear extremal functions.  

In total, we have $2 + 4 + k-1 = k+5$ edges that yield an ordered bipartite graph with a linear extremal function when removed.  
$\Box$
\end{proof}

\begin{theorem}\label{inf min non lin}(Geneson \cite{Geneson})
There are infinitely many minimally non-linear ordered bipartite patterns.  
\end{theorem}

\begin{proof}
The ordered bipartite graph $H_k$ has $3k+5$ edges.   
Of these edges, there are at least $k+5$ that yield an ordered bipartite graph with a linear extremal function when removed.  
So, for each of the ordered bipartite graphs $H_k$, 
	there is a minimally non-linear ordered bipartite graph with somewhere between $k+5$ and $3k+5$ edges.  
We can pick an infinite sequence $1 \leq k_1 < k_2 < k_3, \dots$ 
	such that $\{k_i+5, \dots, 3k_i+5\} \cap \{k_j+5, \dots, 3k_j+5\} = \emptyset$ for $i \neq j$.  
For instance, let $k_i = 4^i$.  
For each of these intervals there must be at least one unique minimally non-linear ordered bipartite graph.  
So, there must be infinitely many minimally non-linear ordered bipartite patterns.  
$\Box$
\end{proof}

\section{A conjecture for the extremal functions of trees}
In \cite{PachTardos}, Pach and Tardos make Conjecture~\ref{tree conjecture} 
	which states that for all pattern graphs $P$ with $\chi_<(P) = 2$ whose underlying graphs are forests, $ex_<(n, P) = O(n \log^{O(1)} n)$.  
Pach and Tardos further conjecture that $ex_<(n, P) = O(ex_<(n, P') \log n)$
    where $P$ is a pattern graph with $\chi_<(P) = 2$ that has a vertex of degree one
    and $P'$ is the pattern obtained from $P$ by removing that vertex. 
Because every tree has a vertex of degree one, this conjecture would imply Conjecture~\ref{tree conjecture}. 

Though Conjecture \ref{tree conjecture} remains open,
    we have a number of similar rules that we may use to unravel our pattern graphs to obtain bounds on their extremal functions: 
\begin{lemma}\label{tree rules}
\begin{itemize}
\item[(a)]
If $P$ is a subgraph of $P'$,
    then $ex_<(n, P) \leq ex_<(n, P')$. 
\item[(b)](F\"uredi and Hajnal \cite{FurediHajnal})
Let $P$ be an ordered bipartite pattern graph with $U(P) = \{u_1 < \dots < u_m\}$.  
If $P'$ is created from $P$ by appending a single vertex $u_{m+1}$ to $U(P)$ such that $u_{m+1} > u_m$ 
	and $u_{m+1}$ is adjacent to exactly one of the neighbors of $u_m$,
    then $ex_2(n, P') \leq ex_2(n, P) + n$. 
\item[(c)](Tardos \cite{Tardos})
If $P'$ is created from $P$ by inserting a single vertex $v$ of degree one
    between two consecutive vertices that are both adjacent to $v$'s neighbor,
    then $ex_<(n, P') \leq 2ex_<(n, P)$. 
\item[(d)](Keszegh \cite{Keszegh})
Let $R = (U_R, V_R, E_R)$ be a graph 
	such that $U_R = \{u_1 < \dots < u_x < \dots < u_s\}$, 
	$V_R = \{v_1 < \dots < v_y < \dots < v_t\}$, 
	and all edges are either between $\{u_1< \dots < u_x\}$ and $\{v_1 < \dots < v_y\}$
	or between $\{u_x, \dots, u_s\}$ and $\{v_y, \dots, v_t\}$.  
Let $P$ be the ordered bipartite graph induced on vertices $\{u_1< \dots < u_x\}$ and $\{v_1 < \dots < v_y\}$, 
	$Q$ be the ordered bipartite graph induced on vertices $\{u_x, \dots, u_s\}$ and $\{v_y, \dots, v_t\}$, 
	and $u_xv_y \in E(R)$.  
Then $ex_2(n, R) \leq ex_2(n, P) + ex_2(n, Q)$. 
\item[(e)](Tardos \cite{Tardos})
If $P'$ is created from $P$ by removing all of the isolated vertices of $P$,
    then $ex_2(n, P) = O(ex_2(n, P') + n)$. 
\item[(f)](Pach, Tardos \cite{PachTardos})
Let $P$ be a pattern graph with vertices $u_0, u_1 \in U(P)$, consecutive vertices $v_0 < v_1 < v_2 \in V(P)$,   
	and edges $u_0v_1, u_0v_2, u_1v_0$, and $u_1v_2$
    and with $u_0v_1$ as the only edge incident on $v_1$. 
Then if $P'$ is created by removing vertex $v_1$ from $P$,
    then $ex_2(n, P) = O(ex_2(n, P') \log n)$. 
\item[(g)](Pach, Tardos \cite{PachTardos})
Let $P$ pattern graph with vertices $u_0, u_1, \in U(P)$ and consecutive vertices $v_0 < v_1 < v_2 < v_3 \in V(P)$
	and edges $u_0v_0$, $u_0v_1$, $u_1v_2$, and $u_1v_3$ with $d(v_1) = d(v_2) = 1$.  
If $P'$ is created from $P$ by removing vertices $v_1$ and $v_2$,
    then $ex_2(n, P) = O(ex_2(n, P') \log^2 n)$. 
\end{itemize}
\end{lemma}

\begin{proof}
Part (a) is trivial for if a graph contains $P'$, then surely it contains $P$. 

For part (b), say that we have a graph $G'$ with $|U(G')| = |V(G')| = n$ that avoids $P'$ and has $ex_2(n, P')$ edges. 
For each non-isolated vertex $u \in U'$, we may remove the edge $\{u, v\}$ where $v = \max\{w : \{u, w\} \in E(G')\}$. 
We remove at most $n$ edges here. 
We are left with a graph $G$ which must avoid $P$ (for otherwise $G'$ would contain $P'$),
    so we have that $|E(G)| = ex_2(n, P') - n \leq ex_2(n, P)$. 

For part (c), say that we have a graph $G'$ with $|V(G')| = n$ that avoids $P'$ and has $ex_<(n, P')$ edges. 
For each vertex of $u \in U'$, order the edges incident on $u$ with $u$ as the smaller vertex and remove every second edge.  
We remove at most half of the edges here.  
We are left with a graph $G$ which must avoid $P$ (for otherwise $G'$ would contain $P'$), 
    so we have that $|E(G)| = 1/2 ex_<(n, P') \leq ex_<(n, P)$.  
	
For part (d), consider a graph $G$ avoiding $R$ with $|U(G)| = |V(G)| = n$ and $|E(G)| = ex_2(n, R)$.  
Let us call an edge ``type one'' if it is incident on the last vertices of each partite set of a subgraph of $G$ isomorphic to $P$ 
	and an edge ``type two'' if it is incident on the first vertices of each partite set of a subgraph of $G$ isomorphic to $Q$.  
Let $G'$ be the graph obtained from $G$ by removing all edges of type one
	and $G''$ be the graph obtained from $G$ by removing all edges of type two.  
Notice that $G' \leq ex_2(n, P)$ and $G'' \leq ex_2(n, Q)$.  
An edge may not be both type one and type two for otherwise $G$ would contain $R$.  
So, we have that $|E(G)| \leq |E(G')| + |E(G'')| \leq ex_2(n, P) + ex_2(n, Q)$.  

For part (e), let $k$ be large enough so that there are fewer than $k$ consecutive isolated vertices in the partite sets of $P$.  
Let $G$ be a graph avoiding $P$ such that $U(G) = \{u_1 < \dots < u_n\}$, $V(G) = \{v_1 < \dots < v_n\}$, and $|E(G)| = ex_2(n, P)$.  
Let us construct a graph $G'$ by removing all edges from the first and last $k$ vertices from each of the partite sets of $G$.  
We remove at most $4kn$ edges here.  
From this graph $G'$, let us define $G'_{i, j}$ for $0 \leq i, j < k$ 
	as $U(G'_{i, j}) = U(G')$, $V(G'_{i, j}) = V(G')$, and $E(G'_{i, j}) = \{u_sv_t : u_sv_t \in E(G'), s = i \mod k, t = j \mod k\}$.  
Each of these $G'_{i, j}$ must avoid $P'$ for otherwise $G$ would contain $P$.  
So, $|E(G'_{i, j})| < ex_2(n, P')$ for all $0 \leq i, j < k$.  
Also, $\sum_{i = 0}^{k-1} \sum_{j=0}^{k-1} |E(G'_{i, j})| = |E(G')|$.  
So, we have $ex_2(n, P) = |E(G)| \leq |E(G')| + 4kn \leq k^2 ex_2(n, P') + 4kn = O(ex_2(n, P') + n)$.  

For part (f), let $G$ be a graph avoiding $P$ with $U(G) = \{u_1 < \dots < u_n\}$, $V(G) = \{v_1 < \dots < v_n\}$, and $|E(G)| = ex_2(n, P)$.  
For $1 \leq i, j \leq n$, let $prev_{ij}$ be defined as the largest $j'$ such that $j' < j$ and $u_iv_{j'} \in E(G)$.  
If there is no such $j'$, then $prev_{ij}$ is undefined.  
For $0 \leq l \leq \lfloor \log n \rfloor$, 
	define graph $G_l$ as $U(G_l) = U(G)$, $V(G_l) = V(G)$, 
	and $E(G_l) = \{u_iv_j : u_iv_j \in E(G), 2^l  \leq j - prev_{ij} < 2^{l+1}\}$.  
Notice that $|E(G)| - n \leq \sum_{l = 0}^{\lfloor \log n \rfloor} |E(G_l)|$.  
Let us define $G'_l$ as the graph obtained from $G_l$ 
	by removing every other edge (with respect to the ordering of $V(G_l)$) from each of the vertices of $u \in U(G_l)$ 
	while keeping $d(u) \geq |E(u)|/2$.  
Note that vertices in $V(G'_l)$ that are both adjacent to the same vertex of $U(G'_l)$ 
	have at least $2^{l+1}$ vertices between them with respect to the ordering of the partite sets.  

Each of the graphs $G'_l$ avoids $P'$.  
Assume to the contrary that $G'_l$ contains $P'$ for some $l$.  
Then there are vertices $u, u' \in U(G'_l)$ that stand for vertices $u_0$ and $u_1$ 
	and vertices $v < v' \in U(G'_l)$ that stand for $v_0$ and $v_2$.  
Since $v$ and $v'$ are at least $2^{l+1}$ apart with respect to the ordering of $V(G)$ and are vertices of $G'_l$, 
	by the definition of $G_l$ there must have been a vertex $v'' \in V(G)$ with edge $u'v''$ that could stand for $v_1$.  
This means that $G$ must contain $P$ which is a contradiction.  
So, we have that $ex_2(n, P) = |E(G)| \leq n + 2\sum_{l = 0}^{\lfloor \log n \rfloor} |E(G_l')| = O(ex_2(n, P')\log n)$.  

For part (g), let $G$ be a graph with $U(G) = \{u_1 < \dots < u_n\}$, $V(G) = \{v_1 < \dots < v_n\}$, and $|E(G)| = ex_2(n, P)$.  
For $1 \leq i, j \leq n$, 
	let $prev_{ij}$ ($next_{ij}$) be defined as the largest (smallest) $j'$ such that $j' < j$ ($j' < j$)and $\{u_i, v_{j'}\} \in E(G)$.  
If there is no such $j'$, then $prev_{ij}$ ($next_{ij}$) is undefined.  
For $0 \leq k, l \leq \lfloor \log n \rfloor$, 
	define graph $G_{kl}$ as $U(G_{kl}) = U(G)$, $V(G_{kl}) = V(G)$, 
	and $E(G_{kl}) = \{u_iv_j : u_iv_j \in E(G), 2^k  \leq j - prev_{ij} < 2^{k+1}, 2^l \leq next_{ij} - j < 2^{l+1}\}$.  
Notice that $|E(G)| - 2n \leq \sum_{k = 0}^{\lfloor \log n \rfloor}\sum_{l = 0}^{\lfloor \log n \rfloor} |E(G_{kl})|$.  
Also, let us define $G_{kl}'$ as the graph obtained from $G_{kl}$ 
	by removing every other edge (with respect to the ordering of $V(G_{kl})$) from each of the vertices of $u \in U(G_{kl})$
	while keeping the degree of the vertex at least $|E(u)|/2$.  
Notice that vertices in $V(G_{kl}')$ that are both adjacent to the same vertex of $U(G_{kl}')$
	have at least $2^k + 2^l$ vertices between them with respect to the ordering of $V(G)$.  

Now, each of the graphs $G'_{kl}$ avoids $P'$ because, as in part (f), if they did not, we would be able to obtain $G$ in $P$.  
So, we have that 
	$ex_2(n, P) 
	= |E(G)| 
	\leq 2n + \sum_{k = 0}^{\lfloor \log n \rfloor}\sum_{l = 0}^{\lfloor \log n \rfloor} |E(G_{kl})| 
	\leq 2n + ex_2(n, P')\log^2n
	= O(ex_2(n, P')\log^2n)$.  
$\Box$
\end{proof}

As stated in the introduction, using Lemma~\ref{tree rules}, 
	Conjecture \ref{tree conjecture} has been proven for all trees with at most five edges. 
We provide a bound of $O(n2^{\sqrt{\log n \log\log n}})$ for one of the two remaining trees with six edges in Theorem~\ref{s_matrix theorem}. 
Though weaker than the bound of the conjecture, 
	this bound is still well below the number of edges allowed in graphs avoiding a pattern with cycles.  

\begin{theorem}\label{cycle theorem}(Erd\H os, Sachs \cite{ErdosSachs})  
Let $C_k$ be the cycle of length $k$.  
The maximum number of edges for a $C_k$-free (unordered) graph with $n$ vertices is $\Omega(n^{1 + 1/(k-1)})$. 
\end{theorem}

\begin{proof}
We prove this theorem with a probabilistic argument. 
Let $G$ be a random graph with $n$ vertices and with edges between vertices with probability $p = n^{(2-k)/(k-1)}/2$. 
Then the expected number of edges in our graph will be 
\begin{eqnarray*} 
	p\binom{n}{2} &=& \frac{n^{(2-k)/(k-1)}\binom{n}{2}}{2} \\
				&<& \frac{n^{(2-k)/(k-1)}n^2}{4} \\
				&=& \frac{n^{k/(k-1)}}{4}
\end{eqnarray*}
    and the expected number of cycles of length $k$ will be 
\begin{eqnarray*}
	\frac{p^kn!}{2(n-k)!k} &=& \frac{2^{-k}n^{(2k-k^2)/(k-1)}n!}{2(n-k)!k} \\
				&<& \frac{2^{-k}n^{(2k-k^2)/(k-1)}n^k}{2k} \\
				&=& \frac{n^{k/(k-1)}}{2^{k+1}k}. 
\end{eqnarray*}
Thus, the expected number of edges not in a cycle will be at least $(1/4 - 2^{-k-1})n^{k/(k-1)}$.  
We may choose a graph with the expected number of edges not in $k$-cycles, 
    remove all of the edges that are involved in $k$-cycles, 
	and be left with a $C_k$-free graph with $\Theta(n^{1 + 1/(k-1)})$ edges. 
$\Box$
\end{proof}

Because it is easier for a host graph to avoid a pattern graph with orderings on the vertices of the graphs,
    the number of edges allowed in an ordered host graph which avoids a particular ordered pattern graph
    is at least as high as the number of edges allowed in an unordered graph avoiding the underlying graph of that ordered pattern graph. 
So, as a corollary to Theorem \ref{cycle theorem}, $ex_<(n, G) = \Omega(n^{k/(k-1)})$ 
	where $G$ is an ordered graph that contains a $k$-cycle in its underlying graph. 
 
\chapter{The extremal functions of generalized matchings}

As noted in the introduction, we call an ordered graph $G$ an $m$-tuple matching
    if $V(G) = \{v_1 < \dots < v_{(m+1)k}\}$, 
	and $E(G) = \{v_jv_{k+i+m(\pi(j)-1)} : i \in \{1, \dots, m\}, j \in \{1, \dots, k\}\}$
    where $\pi:k \rightarrow k$ is a permutation.  

\begin{theorem} \label{geneson mod}
If $P$ is a 2-tuple matching, then $ex_<(n, P) = O(n)$.  
\end{theorem}

The proof of this theorem is a slight modification of Geneson's proof in \cite{Geneson}
	which is in turn a modification of Marcus and Tardos's proof in \cite{MarcusTardos}.  
The general idea is that we will put a recursive bound on the number of edges in a graph that avoids $P$.  
	
For this proof we need the following definitions.  
Let $G$ be an ordered graph which avoids $P$ with $|V(G)| = n$ and assume that $2k^2$ divides $n$. 
We partition the vertices of $G$ into $n/(2k^2)$ intervals $I_1, \dots, I_{n/(2k^2)}$ with $2k^2$ vertices each. 
We define \emph{block} $E(I_i, I_j)$ as the set of edges between vertices of $I_i$ and $I_j$.  
For a block $E(I_i, I_j)$ with $i < j$, we call $I_i$ the \emph{left interval} of the block and $I_j$ the \emph{right interval} of the block.  
We call a block $E(I_i, I_j)$ with $i < j$
    \emph{left-heavy} if there are at least $k$ vertices in $I_i$
    adjacent to vertices in $I_j$ 
	and \emph{right-heavy} if there are at least $2k$ vertices of $I_j$
    adjacent to a single vertex of $I_i$. 
Notice that if a block is neither left- nor right-heavy, it contains at most $(2k-1)(k-1)$ edges. 

\begin{lemma} \label{left-heavy}
The number of edges in left-heavy blocks is at most $4k^4 \cdot n/(2k^2) \cdot (k \binom{2k^2}{k})$.  
\end{lemma}

\begin{proof}
For each left interval $I_i$, 
	there are at most $2k\binom{2k^2}{k}$ right intervals $I_j$ such that $E(I_i, I_j)$ is left-heavy. 
If there were more than $2k\binom{2k^2}{k}$ such intervals $I_j$,
    then by the pigeonhole principle there would exist $k$ vertices in $I_i$
    which would all be adjacent to vertices in each of the $2k$ intervals $I_{j_0}, \dots, I_{j_{2k}}$. 
The edges between these vertices could be used to obtain any 2-tuple matching with $2k$ edges as a subgraph. 
A block may contain at most $4k^4$ edges 
	so the number of edges involving left-heavy blocks is at most $4k^4 \cdot n/(2k^2) \cdot (2k\binom{2k^2}{k})$. 
$\Box$
\end{proof}

\begin{lemma} \label{right-heavy}
The number of edges in right-heavy blocks is at most $4k^4 \cdot n/(2k^2) \cdot (k \binom{2k^2}{2k})$.  
\end{lemma}

\begin{proof}
For each right interval $I_j$, 
	there are at most $k\binom{2k^2}{2k}$ left intervals $I_i$ such that $E(I_i, I_j)$ is right-heavy. 
If there were more than $k\binom{2k^2}{2k}$ such intervals $I_i$,
    then by the pigeonhole principle there would exist $2k$ vertices in $I_j$
    which would all be adjacent to a particular vertex in each of some $k$ intervals $I_{i_0}, \dots, I_{i_k}$. 
Again, the edges between these vertices could be used to obtain any 2-tuple matching with $2k$ edges as a subgraph. 
Thus, the number of edges involving right-heavy blocks is at most $4k^4 \cdot n/(2k^2) \cdot (k\binom{2k^2}{2k})$. 
$\Box$
\end{proof}

Now also define a graph $G'$ which is in a sense a condensed version of $G$. 
We identify each interval $I_i$ in $G$ with a vertex $v_i$ in $G'$
    and have these vertices inherit the ordering of the vertices of $G'$ from $G$. 
There is an edge between two vertices $v_i$ and $v_j$ of $G'$ with $i < j$
    if either (1) $j$ is the lowest index such that $j > i$ and the block $E(I_i, I_j)$ is non-empty
    or (2) there is a single vertex in $I_i$ that is adjacent to at least two vertices 
	in the union of intervals $I_{j'}, I_{j'+1}, \dots, I_{j-1}, I_j$
    where $j'$ is the greatest index such that $j' < j$ and there is an edge between $v_i$ and $v_j$.  
We allow no loops in $G'$.  

\begin{lemma} \label{G' avoids P}
$G'$ avoids $P$.  
\end{lemma}

\begin{proof}
Assume to the contrary that $G'$ contains $P$. 
Then $P$ is a subgraph on some vertices $v_1 < \dots < v_{3k}$ of $G'$.  
This subgraph must have edges $v_iv_{k+2f(i)-2}$ and $v_iv_{k+2f(i)-1}$ 
	for $i \in \{1, \dots, k\}$.  
Since there are edges $v_iv_{k+2f(i)-2}$ and $v_iv_{k+2f(i)-1}$, 
	by the construction of $G'$ there must be a vertex in the interval $I_i$ in $G$ 
	with edges incident on vertices in the intervals $I_{k+2f(i)-2}$ and $I_{k+2f(i)-1}$ for all $i \in \{1, \dots, k\}$.  
These edges form a subgraph of $G$ isomorphic to $P$ which is a contradiction because $G$ avoids $P$.  
$\Box$
\end{proof}

We now group the blocks together.  
For each edge $v_iv_j$, 
	we define a \emph{chunk} as the set of blocks $E(I_i, I_j), \dots, E(I_i, I_{j'-1})$ 
	where $j'$ is the smallest index such that $j' > j$ and $v_iv_j \in E(G')$.  
If there is no such $j'$, then the chunk includes blocks $E(I_i, I_j), \dots, E(I_i, I_{n/(2k^2)})$.  
Notice that each chunk corresponds to one edge in $G'$ and that the chunks partition the non-empty blocks of $G'$.  
We are now ready for the proof of our main lemma.  

\begin{lemma}\label{geneson lemma}
Let $P$ be a 2-tuple matching with permutation $\pi:k \rightarrow k$ and $k \geq 2$. 
If $2k^2$ divides $n$, then $ex_<(n, P) \leq n \cdot 11k^3 \binom{2k^2}{2k} + (2k-1)(k-1)ex_<(n/(2k^2), P)$. 
\end{lemma}

\begin{proof}
Now we bound the total number of edges in $G$.  
First, the total number of intra-interval edges in $G$ is no more than $n/(2k^2)\binom{2k^2}{2}$.  
Those that remain are either in left-heavy blocks, right-heavy blocks, or neither.  
By Lemmas~\ref{left-heavy} and \ref{right-heavy}, 
	the total number of edges in left- or right-heavy blocks 
	is at most $n/(2k^2)(k\binom{2k^2}{2k}) + n/(2k^2)(k\binom{2k^2}{k})$.  

Those edges that remain are nether left- nor right-heavy and are contained in chunks.  
There are two kinds of chunks:  
	(1) those with a single non-empty block 
	and (2) those with more than one non-empty block 
	but with no vertex in the left interval adjacent to more than one vertex in the right intervals for that chunk.  
For each possible left interval, 
	the number of chunks of the second kind 
	with more than $k$ vertices in the left interval adjacent to vertices in the right intervals
	is less than $2k\binom{2k^2}{k}$.  
If there were $2k\binom{2k^2}{k}$ such chunks, 
	then by the pigeonhole principle there would exist $k$ vertices in the left interval
	which would all be adjacent to vertices in the right intervals.  
As before, we could use these vertices to construct any 2-tuple matching with $2k$ edges.  
So, the total number of chunks of the second kind with more than $k$ vertices in the left interval adjacent to vertices in the right interval 
	is less than $n/(2k^2) \cdot 2k \binom{2k^2}{k} = n/k \binom{2k^2}{k}$.  
These contain at most $4k^4 \cdot n/(2k^2) \cdot 2k \binom{2k^2}{k}$ edges.  
The remaining chunks of the second kind contain at most $k$ edges apiece.  

The number of remaining edges in a chunk of the first kind is at most $(2k-1)(k-1)$ edges.  
If the edges are in chunks of the second kind, 
	then either it contains at most $k$ edges or it is in one of the at most $n/k \binom{2k^2}{k}$ chunks with more than $k$ edges 
	(which contain at most $4k^4 \cdot n/k \cdot \binom{2k^2}{k}$ edges).  
By Lemma~\ref{G' avoids P} there are at most $ex_<(n/(2k^2), P)$ chunks total 
	so there are at most $4k^4 \cdot n/k \cdot \binom{2k^2}{k} + (2k-1)(k-1)ex_<(n/(2k^2), P)$ edges 
	between intervals in blocks that are neither left- nor right-heavy.  
So, we can bound the total number of edges in $G$ by
\begin{eqnarray*}
ex_<(n, P) 	&\leq& 	\frac{n}{2k^2}\binom{2k^2}{2}
					+ 4k^4 \frac{n}{2k^2} 2k\binom{2k^2}{k}
					+ 4k^4 \frac{n}{2k^2} k \binom{2k^2}{2k} \\
			&	&	
					+ 4k^4 \frac{n}{k} \binom{2k^2}{k}
					+ (2k-1)(k-1)ex_<(n/(2k^2), P)\\  
			&\leq& 	n \cdot 11k^3 \binom{2k^2}{2k}
					+ (2k-1)(k-1)ex_<(n/(2k^2), P)\\
\end{eqnarray*}
where the last inequality depends on the fact that $k \geq 2$.  
$\Box$
\end{proof}

\noindent{\em Proof of Theorem~\ref{geneson mod}:}
First notice that if $k = 1$, the statement is trivial.  So, assume that $k \geq 2$.  

We will prove that $ex_<(n, P) \leq 11k^4\binom{2k^2}{2k}n$ for all $2$-tuple matchings $P$ by induction on $n$.  
Firstly, notice that this statement is true for $n \leq 2k^2$.  
Now, assume that this statement is true for all $n < m$ and now let $n = m$.  
Also, let $N$ be the largest multiple of $2k^2$ which is less than or equal to $n$.  
Since $k \geq 2$ and $ex_<(n, P) \leq ex_<(N, P) + 2k^2n$, we have by Lemma~\ref{geneson lemma} that
\begin{eqnarray*}
ex_<(n, P) &\leq& ex_<(N, P) + 2k^2n \\
			&\leq& 11k^3\binom{2k^2}{2k}N + (2k-1)(k-1)ex_<(\frac{N}{2k^2}, P) + 2k^2n \\
			&\leq& 11k^3\binom{2k^2}{2k}n + (2k-1)(k-1)11k^4\binom{2k^2}{2k}\frac{n}{2k^2} + 2k^2n \\
			&\leq& k^2\binom{2k^2}{2k}n (11 + (11/2)(2k-1)(k-1) + 2) \\
			&\leq& 11k^4\binom{2k^2}{2k}n
\end{eqnarray*}
$\Box$

\begin{corollary}\label{matching theorem}
If $P$ is a $m$-tuple matching, then $ex_<(n, P) = O(n)$.  
\end{corollary}

\begin{proof}
As in \cite{Geneson}, we can prove this theorem with induction on $m$.  
Theorem~\ref{geneson mod} proves the case for $m = 2$ and $m = 1$.  
Now, assume that the corollary holds for $m < l$ and let $m = l$.  
Let $P$ be an $m$-tuple matching with permutation $\pi:k \rightarrow k$.  
We may apply Lemma~\ref{tree rules}(c) $k$ times to remove $k$ edges and obtain an $(m-1)$-tuple matching $P'$ with the same permutation.  
In removing these edges we pay a penalty of a factor of $2^k$.  
So, we have $ex_<(n, P) \leq 2^k ex_<(n, P') = O(n)$.  
$\Box$
\end{proof}

Now we may go back and prove Theorem~\ref{geneson main} concerning the extremal function of $m$-tuple bipartite matchings.  \\

\noindent\emph{Proof of Theorem~\ref{geneson main}}
Let $P$ be a $m$-tuple bipartite matching.  
From an ordered bipartite graph $G$ that avoids $P$ 
	we can construct an ordered (non-bipartite) graph $G'$
	that also avoids $P$
	with the same number of edges.  
We can do this by letting $V(G') = U(G) \cup V(G)$ where all of the vertices of $V(G)$ follow those of $U(G)$ 
	and carrying over all of the edges.  
So, since $ex_<(n, P) = O(n)$, $ex_2(n, P) = O(n)$ for every $m$-tuple bipartite matching $P$.  
$\Box$

Corollary~\ref{matching theorem} also implies a generalization 
	of a result for convex geometric graphs originally proven in \cite{AFPS}.  
Their result proved that the $ex_\circlearrowright$ is linear for convex geometric matchings.  
We call a convex geometric graph $G$ an \emph{$m$-tuple convex geometric matching} 
	if the underlying graph of $G$ is isomorphic to the underlying graph of a $m$-tuple matching,  
	$\chi_\circlearrowright(G) = 2$, 
	and all of the vertices of degree one form an interval.  

\begin{corollary}\label{tuple circular}
If $P$ is a $m$-tuple convex geometric matching, then $ex_\circlearrowright(n, P) = O(n)$.  
\end{corollary}

\begin{proof}
Let $P$ be a $m$-tuple convex geometric matching 
	and let $P'$ be the $m$-tuple matching defined by using the same permutation as $P$.  
As in the proof of \ref{geneson main}, 
	we can use any convex geometric graph $G$ that avoids $P$
	to construct a ordered graph $G'$ that avoids $P'$.  
We do this by simply changing the ordering of the vertices of $G$ 
	from cyclic to linear (with an arbitrary initial vertex).  
Again, since $ex_<(n, P') = O(n)$ for $m$-tuple matchings $P'$, 
	$ex_\circlearrowright(n, P) = O(n)$ for $m$-tuple convex geometric matchings.  
$\Box$
\end{proof}

Corollary~\ref{tuple circular} was proven previously for the case $k = 1$ by Eyal Ackerman, Jacob Fox, J\'anos Pach, Andrew Suk \cite{AFPS} 
	and independently proven for the general case by Bal\'azs Keszegh \cite{KeszeghPersonal}.   
\chapter{The extremal function of the sailboat graph}\label{chapter s_matrix}

This work also appears in \cite{TardosWeidert}.

\begin{theorem}\label{s_matrix theorem}
For the ``sailboat graph'' pattern $P$ in Figure~\ref{sailboat} we have
$$ex_2(n,P)=n2^{O(\sqrt{\log n \log\log n})}.$$  
\end{theorem}

We start with a simple process that makes an ordered bipartite graph
halfway regular without losing too many edges or introducing a subgraph
isomorphic to $P$. We need this regularization for technical reasons.

For an ordered bipartite graph $G = (U,<_U, V,<_V, E)$ and a parameter $q \geq 1$
	we construct another ordered bipartite graph $G^{(q)}=(U',<_{U'},V,<_V, E')$ as follows. 
We obtain $G^{(q)}$ by ``splitting'' every vertex $u\in U$ into degree $q$ vertices. 
Formally $U'$ is obtained from $U$ by replacing every vertex $u \in U$ by $\lfloor d_G(u)/q \rfloor$ vertices.
The linear ordering $<_{U'}$ is inherited from the ordering of $U$ 
	when comparing two vertices coming from different vertices of $U$ 
	and is arbitrary when comparing vertices of $U'$ obtained from the same vertex of $U$. 
The edge set $E'$ is chosen such that each vertex $u' \in U'$ obtained from a vertex $u \in U$ 
	has exactly $q$ neighbors in $V$ among all the neighbors of $u$ 
	such that distinct vertices $u',u'' \in U'$ coming from the same vertex $u \in U$ have disjoint sets of neighbors. 
Notice that $G^{(q)}$ is typically not uniquely defined.

\begin{lemma}
\label{preprocess avoidance}
Let $G = (U,<_U, V,<_V, E)$ an ordered bipartite graph and $q\ge1$. The degree
of every vertex $u'\in U(G^{(q)})$ is $d(u')=q$ and $G^{(q)}$ has at least
$|E|-(q-1)|U|$ edges. Furthermore if $G$ avoids a pattern $T$ and any two
consecutive vertices in $U(T)$ have a common neighbor in $T$,
then $G^{(q)}$ also avoids $T$.
\end{lemma}

\noindent{\em Proof:} The regularity follows from the construction. The bound
on the edges follows from the fact that we loose at most $q-1$ edges incident
to any vertex $u\in U$.

For the last statement we prove the contrapositive. Assume that $G'=G^{(q)}$
contains $T$. Let $T'$ be subgraph of $G'$ isomorphic to $T$ and let $u_1' <
\dots < u_x'$ be the vertices of $U(T')$. Let $u_i\in U$ be the vertex $u_i'$
is created from. We have $u_1\le\ldots\le u_x$ and since each consecutive pair
$u_i'$ and $u_{i+1}'$ share a common neighbor we must have $u_i\ne
u_{i+1}$. Thus we have $u_1<\ldots<u_x$. Replacing the vertices $u_i'$ in $T'$
by $u_i$ we get an isomorphic copy of $T$ as a subgraph of $G$ finishing the
proof.
$\Box$

\begin{lemma}
\label{preprocess lemma}
For any $n$ there is an ordered bipartite graph $G$ avoiding $P$ with $|U(G)| =
|V(G)| = n$ and $d_G(u)=\lfloor ex_2(n,P)/(2n)\rfloor$ for every $u\in U(G)$.
\end{lemma}

\noindent{\em Proof:}
We start with the extremal ordered bipartite graph $G_n$ avoiding $P$ and
having $|U(G)|=|V(G)|=n$ and $|E(G)|=ex_2(n,P)$. Let $q = \lfloor
ex_2(n,P)/(2n) \rfloor$. In case $q=0$ the statement of the lemma trivially
holds (the claimed ordered bipartite graph $G$ has no edges). We may therefore
assume $q\ge1$ and consider $G'=G_n^{(q)}$. By Lemma~\ref{preprocess
avoidance} $G'$ has at least $ex_2(n,P)-n(q-1)>nq$ edges and each vertex in
$U(G')$ have degree $q$. Thus $|U(G')|>n$. We obtain $G$ from $G'$ as an
induced subgraph by removing
the excess vertices from $U(G')$ and keeping only $n$ of them. We have
$V(G)=V(G')=V(G_n)$, so $|V(G)|=n$. Finally, since both pairs of consecutive
vertices in $U(P)$ has a common neighbor $G'$ avoids $P$ by
Lemma~\ref{preprocess avoidance} and so $G$, being a subgraph of $G'$, must
also avoid it.
$\Box$

The core of the proof of Theorem~\ref{s_matrix theorem} lies in the following
lemma. It claims that given a reasonably dense ordered bipartite graph
avoiding $P$ we can construct a smaller but even denser ordered bipartite
graph still avoiding $P$. The proof of the theorem is then the simple
observation that if the first ordered bipartite graph is somewhat dense then
after a few recursive applications of this lemma we get a graph with
impossibly high density.

\begin{lemma}
\label{main lemma}
Given an ordered bipartite graph $G$ avoiding $P$
with $|U(G)| = |V(G)| = n\ge2$ and $d(u) = t>24\lceil\log n\rceil$ for all $u
\in U$ there exist another ordered bipartite graph $G'$ avoiding $P$ with
$|U(G')| = |V(G')| = \lfloor n/t \rfloor$ and $d(u)=t'$ for all $u \in U(G')$
with $t'>t/(80\log n)$.
\end{lemma}

\noindent{\em Proof:}
We write $U=U(G)$ and we use $<$ to denote its ordering in $G$. We also write
$E=E(G)$ and identify $V=V(G)$ with $\{1, \dots, n\}$ respecting its
ordering.
We partition $V$ into $t$ intervals $V_1,\ldots,V_t$ of size $n'=\lfloor
n/t\rfloor$ or $n'+1$ each, with the vertices in $V_i$ preceding the vertices
in $V_j$ if $i<j$. We set $p=\left\lfloor t-1\over12\lceil\log n\rceil\right
\rfloor\ge2$. For each vertex $u\in U$ and each interval $V_i$ we partition
the edges of $E$ connecting $u$ to $V_i$ into sets of size $p$ each, with a
possible leftover less than $p$ edges. We call these parts of size $p$ {\em
stars} and all edges outside the stars {\em leftover edges}.

We finish the proof of the lemma assuming there are at least $n$ stars. We
will justify this assumption later. If there are at least $n$ stars, then
there must exist an interval $V_i$ such that at least $n/t\ge n'$ stars
consist of edges incident to $V_i$. Consider such an interval and the subgraph
$G_i$ of $G$ induced by the vertices $U$ and $V_i$. Each of these stars
contribute a vertex to the graph $G_i^{(p)}$, so $|U(G_i^{(p)})|\ge n'$. Now
let $G'$ be the subgraph of $G_i^{(p)}$ obtained by keeping $n'$ of the
vertices $U(G_i^{(p)})$ and $V(G_i^{(p)})=V_i$ each, and exactly $t'=p-1$ edges
incident to all vertices in $U(G')$. This is possible as the vertices in
$U(G_i^{(p)})$ have degree $p$ and at most one of the vertices of
$V(G_i^{(p)})$ gets removed. The graph $G'$ satisfies the size and
degree conditions in the lemma. To see that it avoids $P$ notice that $G_i$
being a subgraph of $G$  avoids $P$ and by Lemma~\ref{preprocess avoidance}
this property is inherited by $G_i^{(p)}$ of which $G'$ is a subgraph.

It remains to prove that we have at least $n$ stars. For this we assume that
there are fewer and reach a contradiction.

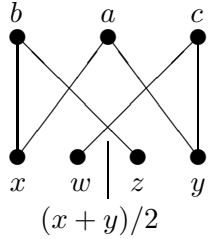
\begin{figure}[h!]
\centering
\begin{picture}(16, 15.5)
\multiput(2,12.75)(6,0){3}{\circle*{1}} 
\multiput(2,4.75)(4,0){4}{\circle*{1}} 
\put(2,4.75){\line(0,1){8}} 
\put(2,4.75){\line(3,4){6}} 
\put(6,4.75){\line(1,1){8}} 
\put(8, 2){\line(0, 1){3.75}}
\put(10,4.75){\line(-1,1){8}} 
\put(14,4.75){\line(-3,4){6}} 
\put(14,4.75){\line(0,1){8}} 

\put(1.5, 13.75){$b$}
\put(7.5, 13.75){$a$}
\put(13.5, 13.75){$c$}
\put(1.5, 2.5){$x$}
\put(5.5, 2.5){$w$}
\put(9.5, 2.5){$z$}
\put(13.5, 2.5){$y$}
\put(3.5,0){$(x+y)/2$}
\end{picture}
\caption{A hat with both a left and a right extension forms the pattern $P$.}
\label{arm figure}
\end{figure}

We call a pair of edges $(ax,ay)$ from $E$ a {\em hat} if they share
a vertex $a\in U$ and satisfy $x<y$. The {\em width} of the hat is
$y-x$. We say that a hat $(bx,bz)$ is a {\em left extension} of the hat
$(ax,ay)$ if $b<a$ and its width satisfies $(y-x)/2<z-x<y-x$. The hat
$(cw,cy)$ is a {\em right extension} of the hat $(ax,ay)$ if $a<c$ and its
width satisfies the same inequalities: $(y-x)/2<y-w<y-x$. If a hat would have
both a left and a right extension then the six involved edges would form a
subgraph isomorphic to $P$, a contradiction. See Figure~\ref{arm figure} for
an illustration. There are $n{t\choose2}$ hats in total, so by symmetry
(notice that $P$ is symmetric too) we may assume that at least
$n{t\choose2}/2$ of them has no right extension. Let us choose an integer $k$
and call the hat $(ax,ay)$ {\em good} if it has no right extension and its
width satisfies $k\le y-x<2k$. As every hat without a right extension is good
for one of the values $k=2^i$ for $i=0,1,\ldots,\lceil\log n\rceil-1$ we can
and will choose $k$ in such a way that there are at least
$n{t\choose2}/(2\lceil\log n\rceil)$ good hats.

We divide the good hats into one of three types:   
\begin{itemize}
\item \textbf{Type One.}
This type contains all good hats $(ax,ay)$ with the edge $ax$ contained in a
star. By our assumption, there are fewer than $n$ stars. Each of these stars
contain $p$ edges and each of these edges are involved in exactly $t-1$
hats. So there are fewer than $npt$ good hats of this type. 

\item \textbf{Type Two.}
This type contains all good hats $(ax,ay)$ not in the first type with $x$ in
one of the intervals $V_i$ satisfying that no good hat $(az,ay)$ exists with
$z\in V_j$ for some $j>i$. Each of the $nt$ edges $ay\in E$ may pair up only
with edges $ax$ with $x$ from a specific single interval $V_i$ to form a good
hat of this type. As $ax$ must also be a leftover edge there are fewer than $p$
choices for any fixed edge $ay$, so the total number of good hats of this type
is less than $npt$.

\item \textbf{Type Three.}  
This type contains all other good hats.  
For each vertex $y\in V$ and interval $V_i$ there exist no more than a single
vertex $a\in U$ such that a good hat $(ax,ay)$ of this type exists with $x\in
V_i$. Indeed, assume to the contrary that there exist two vertices $a<c\in U$
and good hats $(ax,ay)$, $(cz,cy)$ of type three with $x,z\in V_i$. As
$(cz,cy)$ is of type three we must have another good hat $(cw,cy)$ with $w\in
V_j$ for some $j>i$. Thus $w>x$ must hold. As both $(ax,ay)$ and $(cw,cy)$
are good hats their width lies between $k$ and $2k$ and so $(y-x)/2<y-w<y-x$
is satisfied making $(cw,cy)$ a right extension of the good hat $(ax,ay)$, a
contradiction. Now using that the choice of $y$ ($n$ possibilities) and the
interval $V_i$ ($t$ possibilities) determine $a\in U$ for a good hat $(ax,ay)$
of type three with $x\in V_i$ we conclude that are fewer than $npt$ good hats
of this type. Indeed, given $a$ and $V_i$ there fewer than $p$ leftover edges
$ax$ with $x\in V_i$.
\end{itemize}

In total, we have fewer than $3npt$ good hats, contradicting our choice of
$p$ and the assumption that the number of good hats is at least
$n{t\choose2}/(2\lceil\log n\rceil)$. This finishes the proof of the lemma.
$\Box$
\smallskip

\noindent{\em Proof of Theorem~\ref{s_matrix theorem}:}
Let us fix $n$ and start with the ordered bipartite graph $G_0$ whose
existence is claimed in Lemma~\ref{preprocess lemma}. $G_0$ avoids $P$ and we
have $|U(G_0)|=|V(G_0)|=n_0=n$ and $d_{G_0}(u)=t_0=\lfloor
ex_2(n,P)/(2n)\rfloor$. We apply Lemma~\ref{main lemma} iteratively to construct
graphs $G_1, G_2, \dots$ of ever-escalating density. $G_i$ is obtained from
$G_{i-1}$ using Lemma \ref{main lemma}, it avoids $P$ and has
$|U(G_i)|=|V(G_i)|=n_i$ and satisfies $d_{G_i}(u)=t_i$ for each $u\in U(G_i)$
with 
$$n_i=\left\lfloor n_{i-1}\over t_{i-1}\right\rfloor,$$
$$t_i>{t_{i-1}\over80\log n_{i-1}}.$$
We can continue this graph sequence till the
condition $t_i>24\lceil\log n_i\rceil$ is satisfied.
 
The sequence $n_i$ is non-increasing and we can assume the same about $t_i$,
so we have
$$t_i > \frac{t_0}{80^i\log^i n},$$ 
	$$n_i < \frac{n80^{\binom{i}{2}}\log^{\binom{i}{2}}n}{t_0^i},$$  
	and 
	$$\frac{t_i}{n_i} > \frac{t_0^{i+1}}{n80^{\binom{i+1}{2}}\log^{\binom{i+1}{2}} n}.$$
Since this last quantity is the edge density of $G_i$ it is bounded by $1$ we
have for all $i$ 
	\begin{eqnarray*}
	t_0 &<& [n80^{\binom{i+1}{2}}\log^{\binom{i+1}{2}}n]^{1/(i+1)} \\ 
		&=& n^{1/(i+1}80^{i/2}\log^{i/2}n.   
	\end{eqnarray*}
We let $i_0 = \lfloor\sqrt{\log n / \log\log n}\rfloor$ and obtain
$$t_0<80^{i_0/2}2^{1.5\sqrt{\log n \log\log n}}.$$
It is easy to verify that starting with $t_0$ above this threshold the graph
sequence $G_i$ will not stop before $G_{i_0}$ is reached, so the above
inequality must hold. Using $t_0=\lfloor ex_2(n,P)/(2n)\rfloor$ the statement of
the theorem follows.
$\Box$

 
\chapter{Conclusion}

In this paper we bound the extremal function of $m$-tuple matchings in ordered graphs 
	and make some incremental progress in characterizing $ex_2(n, P)$ where $P$ is a tree.  
The extremal function $ex_2(n, P)$ for the ``sailboat'' ordered bipartite graph $P$ of Figure~\ref{sailboat} 
	lacks a tight lower bound.  
If indeed the upper bound is tight, this graph would provide a counter-example to Conjecture~\ref{tree conjecture}.  
Also, it would be interesting to see whether the techniques used in Chapter~\ref{chapter s_matrix}
	may be applied to the extremal functions for other fixed graphs besides the one in Figure~\ref{sailboat}.  
The extremal functions of one more tree on six edges (explicitly given in \cite{PachTardos}) 
	and many trees of more than six edges are yet unknown.



\end{document}